
\input harvmac
\def\listrefssave
{\footatend\bigskip\bigskip\bigskip\immediate\closeout\rfile\writestoppt
\baselineskip=14pt\centerline{{\bf References}}\bigskip{\frenchspacing%
\parindent=20pt\escapechar=` \input \jobname.refs\vfill\eject}\nonfrenchspacing}

\def\CTPa{\it Center for Theoretical Physics, Department of Physics,
      Texas A\&M University}
\def\CTPb{\it College Station, TX 77843-4242, USA}

\def\HARCa{\it Astroparticle Physics Group,
Houston Advanced Research Center (HARC)}
\def\HARCb{\it The Woodlands, TX 77381, USA}

\def\ie{\hbox{\it i.e.}}     
\def\eg{\hbox{\it e.g.}}

\catcode`\@=11 

\def\lsim{\mathrel{\mathpalette\@versim<}}
\def\gsim{\mathrel{\mathpalette\@versim>}}
\def\@versim#1#2{\vcenter{\offinterlineskip
    \ialign{$\m@th#1\hfil##\hfil$\crcr#2\crcr\sim\crcr } }}
\def\boxit#1{\vbox{\hrule\hbox{\vrule\kern3pt
      \vbox{\kern3pt#1\kern3pt}\kern3pt\vrule}\hrule}}

\def\etal{{\it et. al.}}

\def\t1{{\tilde 1}}
\def\ov{\overline}

\def\JL{J. L. Lopez}
\def\DVN{D. V. Nanopoulos}

\def\NPB#1#2#3{Nucl. Phys. B {\bf#1} (19#2) #3}
\def\PLB#1#2#3{Phys. Lett. B {\bf#1} (19#2) #3}

\def\PRD#1#2#3{Phys. Rev. D {\bf#1} (19#2) #3}
\def\PRL#1#2#3{Phys. Rev. Lett. {\bf#1} (19#2) #3}
\def\PRT#1#2#3{Phys. Rep. {\bf#1} (19#2) #3}

\def\IJMP#1#2#3{Int. J. Mod. Phys. A {\bf#1} (19#2) #3}
\def\TAMU#1{Texas A \& M University preprint CTP-TAMU-#1}

\nref\flipr{For a recent review see \eg, \JL\ and \DVN, \TAMU{76/91},
to appear in Proceedings of the 15th Johns Hopkins Workshop on Current Problems
in Particle Theory, August 1991.}
\nref\Der{J. P. Derendinger, L. Ib\'a\~nez, and H. Nilles, \PLB{155}{85}{65};
M. Dine, R. Rohm, N. Seiberg, and E. Witten, \PLB{156}{85}{55}.}
\nref\ssb{
I. Antoniadis, J. Ellis, A. B. Lahanas, and \DVN, \PLB{241}{90}{24};
L. Dixon, in Proceedings of The Rice Meeting, ed. by B.
Bonner and H. Miettinen (World Scientific, 1990), p. 811;
C. P. Burguess and F. Quevedo, \PRL{64}{90}{2611};
A. Font, L. Iba\~nez, D. L\"ust, and F. Quevedo, \PLB{245}{90}{401};
S. Ferrara, N. Magnoli,
T. R. Taylor, and G. Veneziano, \PLB{245}{90}{409}; H. P. Nilles and
 M. Olechowsky, \PLB{248}{90}{268}; P. Binetruy and M. K. Gaillard,
\PLB{253}{91}{119}; M. Cvetic, \etal, \NPB{361}{91}{194}.}
\nref\revamp{I. Antoniadis, J. Ellis, J. Hagelin, and \DVN, \PLB{231}{89}{65}.}
\nref\decisive{J. L. Lopez and \DVN, \PLB{251}{90}{73}.}
\nref\sharp{\JL\ and \DVN, \PLB{268}{91}{359}.}
\nref\FFF{I. Antoniadis, C. Bachas, and C. Kounnas, Nucl. Phys. B
{\bf 289} (1987) 87; I. Antoniadis and C. Bachas, Nucl. Phys. B {\bf298} (1988)
586; H. Kawai, D.C. Lewellen, and S.H.-H. Tye, Phys. Rev. Lett. {\bf57} (1986)
1832; Phys. Rev. D {\bf34} (1986) 3794; Nucl. Phys. B {\bf288} (1987) 1;
R. Bluhm, L. Dolan, and P. Goddard, Nucl. Phys. B {\bf309} (1988) 330;
H. Dreiner, J. L. Lopez, D. V. Nanopoulos, and D. Reiss, Nucl. Phys. B
{\bf 320} (1989) 401.}
\nref\KLN{S. Kalara, J. Lopez, and \DVN, \PLB{245}{90}{421},
\NPB{353}{91}{650}.}
\nref\SK{S. Kalara, in Proceedings of the Superstring Workshop `Strings 90',
ed. by R. Arnowitt, \etal\ (World Scientific, 1991), p. 461 and references
therein.}
\nref\Schwarz{See \eg, J. Schwarz, Caltech preprint CALT-68-1581 (1989).}
\nref\Ferr{S. Ferrara, D. L\"ust, A. Shapere, and S. Theisen,
\PLB{225}{89}{363}; S. Ferrara, D. L\"ust, and S. Theisen, \PLB{233}{89}{147}.}
\nref\Lauer{J. Lauer, J. Mas, and H. P. Nilles, \PLB{226}{89}{251};
W. Lerche, D. L\"ust, and N. P.Warner, \PLB{231}{89}{417};
P. Candelas, X. De La Ossa, P. S. Green, and L. Parks, \NPB{359}{91}{21}
and \PLB{258}{91}{118}.}
\nref\Porrati{A. Giveon and M. Porrati, \PLB{246}{90}{54} and
\NPB{355}{91}{422}.}
\nref\WI{M. B. Green, J. H. Schwarz, and E. Witten, {\it Superstring Theory},
(Cambridge Univ. Press, 1987).}
\nref\anomaly{G. Lopes Cardoso and B. Ovrut, University of Pennsylvania
preprint
UPR-0464T (1991); J. Louis, SLAC preprint SLAC-PUB-5527 (1991);
J. Derendinger, S. Ferrara, C. Kounnas, and F. Zwirner,
CERN preprints CERN-TH.6004/91 and 6202/91.}
\nref\Jan{See \eg, J. Louis, SLAC preprint SLAC-PUB-5645 (1991) and references
therein.}
\nref\LN{A. Lahanas and \DVN, \PRT{145}{87}{1}.}
\nref\twotwo{M. Cvetic, \PRL{59}{87}{1795}, \PRD{37}{88}{2366}.}
\nref\Dixon{L. Dixon, D. Friedan, E. Martinec, and S. Shenker,
\NPB{282}{87}{13}; S. Hamidi and C. Vafa, \NPB{279}{87}{465}.}
\nref\cys{E. Witten, \NPB{268}{86}{79}; M. Dine and N. Seiberg,
\PRL{57}{86}{2625}; M. Dine and C. Lee, \PLB{203}{88}{371}.}
\nref\NR{E. Witten, \NPB{268}{86}{79}; M. Dine and N. Seiberg,
\PRL{57}{86}{2625}.}
\nref\IB{L. Ib\'a\~nez, W. Lerche, D. L\"ust, S. Theisen, \NPB{352}{91}{435}.}
\nref\DKL{L. Dixon, V. Kaplunovsky, and J. Louis, \NPB{329}{90}{27}.}
\nref\thresh{S. Kalara, \JL, and \DVN, \PLB{269}{91}{84}.}
\nref\EKN{J. Ellis, C. Kounnas, and \DVN, \NPB{247}{84}{373}.}
\nref\Cvetic{M. Cvetic, \etal, in Ref. \ssb}
\nref\STAB{J. L. Lopez and \DVN, \PLB{256}{91}{150}.}
\nref\DSW{M. Dine, N. Seiberg, and E. Witten, \NPB{289}{87}{589};
M. Dine, I. Ichinose, and N. Seiberg, \NPB{293}{87}{253};
J. Atick, L. Dixon, and A. Sen, \NPB{292}{87}{109};
M. Yamaguchi, H. Yamamoto, and T. Onogi, \NPB{327}{89}{704};
M. Yamaguchi, T. Onogi, and I. Ichinose, \IJMP{5}{90}{479};
M. Dine and C. Lee, \NPB{336}{90}{317}.}
\nref\conden{S. Kalara, \JL, and \DVN, \TAMU{69/91}
(to appear in Phys. Lett. B).}
\nref\Kap{V. Kaplunovsky, \NPB{307}{88}{145}; L. Dixon, V. Kaplunovsky, and
J. Louis, \NPB{355}{91}{649}.}
\nref\Lacaze{I. Antoniadis, J. Ellis, R. Lacaze, and \DVN, \PLB{268}{91}{188}.}

\leftline{\titlefont TEXAS A\&M UNIVERSITY}
\leftline{\bf CENTER FOR THEORETICAL PHYSICS}
\Title{\vbox{\baselineskip12pt\hbox{CTP--TAMU--94/91}\hbox{ACT--54}}}
{Modular Invariance and Nonrenormalizable Interactions}
\centerline{S. KALARA, JORGE~L.~LOPEZ,\footnote*{Supported
by an ICSC--World Laboratory Scholarship.} and D.~V.~NANOPOULOS}
\bigskip
\centerline{\CTPa}
\centerline{\CTPb}
\centerline{and}
\centerline{\HARCa}
\centerline{\HARCb}
\vskip .3in
\centerline{ABSTRACT}
We examine the modular properties of nonrenormalizable superpotential terms
in string theory and show that the requirement of modular invariance
necessitates the nonvanishing of certain Nth order nonrenormalizable terms.
In a class of models (free fermionic formulation) we explicitly verify that
the nontrivial structure imposed by the modular invariance is indeed present.
Alternatively, we argue that after proper field redefinition, nonrenormalizable
terms can be recast as to display their invariance under the modular group.
We also discuss the phenomenological implications of the above observations.
\bigskip
\Date{November, 1991}

\newsec{Introduction}
It is one thing to believe that string theory is the Theory of Everything
and another to {\it prove} it. The plethora and multiple latitude of problems
that one encounters is well-known: selection of a specific classical vacuum,
stability of the selected vacuum under non-perturbative string effects,
supersymmetry breaking, connection with the experimental reality of low-energy
physics ($E\ll M_{Pl}$), etc. While nobody can claim a panacea for all these
problems, remarkable progress has been made on several fronts. For instance,
(semi) realistic vacua have been identified \flipr\ with no obvious
or presently identifiable stability problems, and different realistic scenarios
have been proposed for ways to break supersymmetry in string theory
and to communicate this seed of supersymmetry breaking to the low-energy
spectrum \refs{\Der,\ssb}.

Some time ago we started to explore ways of connecting the aloof string theory
at the Planck scale with the low-energy physics world
\refs{\revamp,\decisive,\sharp}. We found that one
of the most convenient frameworks to address such issues is the so-called
free fermionic formulation of superstrings in four dimensions \FFF. Within this
formulation we elucidated the methods to be used in calculating superpotential
terms at the cubic level \refs{\revamp,\KLN} as well as at arbitrarily high
orders \KLN.
The importance of this program should be clear, since it is by the inclusion of
nonrenormalizable terms (which arise by integrating out massive string modes)
that we expect to obtain a realistic fermion mass spectrum, quark mixing,
proton stability, etc \SK. This program has been decisive in the identification
of realistic models \decisive.

String theory is more subtle than regular field theory. It contains symmetries
beyond the usual gauge symmetries that impose tight constraints on the
allowed effective action. Indeed, it has been found that string duality
or target space modular invariance \refs{\Schwarz,\Ferr,\Lauer} imposes
considerable restrictions on the four-dimensional effective action \Porrati.
In this paper we determine the modular
invariant properties of the cubic and higher-order superpotential in models
built within the free fermionic formulation. We then use these facts to obtain
powerful new results about the $T$-dependence of the calculated superpotential
couplings. These results may have far-reaching phenomenological consequences.

The task of constructing a string theory effective action is marred by a
plethora of subtle and not so subtle problems. If one were to begin with the
standard supergravity action obtained from string \WI(either by dimensional
reduction of by the sigma model approach),
\eqn\I{{\cal G}=K(T,\ov T,m,m^\dagger)+\ln W(m,T)+\ln \ov W(m^\dagger,\ov T),}
where $T$ are the moduli fields, $m$ generically denotes the matter fields,
and $K$ is the K\"ahler function given by
\eqn\II{K(T,\ov T,m,m^\dagger)=-3\ln(T+\ov T)+f(T,\ov T)mm^\dagger,}
it would naively seem that the effective action for the moduli fields obtained
through $\cal G$ is invariant under the modular transformations of the $T$
fields, since these are part of the K\"ahler transformations, that is
\eqna\III
$$\eqalignno{T&\to {aT-ib\over icT+d},\quad ad-bc\in{\bf Z},&\III a\cr
K&\to-3\ln(T+\ov T)+3\ln(icT+d)+3\ln(-ic\ov T+d),&\III b\cr
W&\to (icT+d)^{-3}W.&\III c}$$
The above transformation properties are contained in $T\to g(T)$, where $g$
is an arbitrary function, and $K\to K+F(T)+\ov F(\ov T)$.
However, as it can be easily shown, the transformations in Eqs. \III{} have
an anomaly at the one-loop level \anomaly. For a consistent
treatment of the modular transformation, this anomaly must be incorporated into
the effective action. Furthermore, the inclusion of nonperturbative effects
of a strongly interacting gauge theory requires additional modifications of
the effective action \ssb. In fact, the focus of the efforts involving modular
invariance and effective string theory actions has been directed towards
these modifications \Jan.
However, a tacit assumption made in the above procedure is to presume that
the superpotential $W$ is indeed such that it transforms covariantly as
in Eq. \III{c}, which may or may not be the case.

In a supergravity theory \LN, one may {\it ``choose"} the superpotential such
that it is already endowed with covariant properties under modular
(or K\"ahler) transformations. However, in string theory one does not have
that luxury. For a large class of models, the superpotential can be
explicitly and unambigously calculated \KLN. Hence for the superstring
effective
action to be invariant under modular transformations, the explicit
calculation of the superpotential must exhibit the covariant properties which
are by no means obvious. Additionally, as it has been noted before, to go
from a `generic' string theory model to a specific phenomenologically
interesting one, it is necessary to explore the superpotential $W$ in much
greater depth \refs{\KLN,\SK,\decisive}. Important physical implications of the
model such as the
scale of gauge symmetry breaking and the fermion mass matrices, can only be
extracted after the superpotential is obtained in great detail. Again, the
modular invariance properties of $W$ are a crucial ingredient in the analysis.
\newsec{General remarks}
As a starting point, let us consider the superpotential at the trilinear
level only
\eqn\V{W_3=c_{ijk}m_im_jm_k.}
It is then clear that a trivial choice of modular weights ($-1$) for each
matter field $m_i$, combined with the cubic nature of the superpotential
leads to a modular covariant expression
\eqna\Vi
$$\eqalignno{m_i&\to (icT+d)^{-1}m_i,&\Vi a\cr
W_3&\to(icT+d)^{-3}W_3.&\Vi b\cr}$$
Hence, we see that the strong constraints that modular transformations
might lead to cannot be uncovered until one considers higher-order
(nonrenormalizable) terms in the superpotential.

Over the past few years there has been considerable progress in evaluating
nonrenormalizable terms in a large class of string theory models. Notable
among these are (2,2) symmetric orbifolds \refs{\twotwo,\Dixon}, Calabi-Yau
manifolds \cys, asymmetric orbifolds \twotwo, and free fermionic
formulations \KLN. In all except free
fermionic models, nonrenormalizable terms cannot be unambigously calculated
due to nonperturbative instanton corrections \SK. For the free fermionic case
there exists a powerful machinery that allows one to evaluate explicitly
nonrenormalizable terms up to very high orders \KLN. In addition, the results
thusly obtained enjoy strong nonrenormalization theorems \NR, making them
especially suited to study the interplay of modular invariance and $W$.

Succintly put, the method employed to evaluate a typical nonrenormalizable
term relies on the calculation of the S-matrix elements between different
fields of interest. Starting with a vertex operator (which generates
properly normalized one-particle states), the string sigma model Lagrangian
can be used to calculate the correlators among different fields which
are then used to extract the possible nonrenormalizable terms that may be
present. The key point to note here is that these vertex operators generate
states
which are already normalized, whereas the states in the supergravity action
carry a nontrivial $T$ dependence \refs{\IB,\Ferr}. This is an important
distinction whose true
significance will soon become clear. To avoid confusion we will denote the
matter fields in the string (supergravity) basis with (un)primed fields.

For theories possessing (2,2) worldsheet supersymmetry the identification
of the moduli fields is reasonably straightforward \DKL. However, for theories
with only (2,0) supersymmetry (\eg, free fermionic formulations and
asymmetric orbifolds) no simple, full-proof method exists for this purpose.
In what follows we will consider a scalar massless field with zero potential
as a possible candidate for a modulus field. The modulus field $\Phi$ is
generated by a vertex operator
\eqn\Vii{V_{\Phi(-1)}(z,\bar z)=e^{-c}G_L(z)G_R(\bar z)e^{ik\cdot x},}
where $G_L(z)$ and $G_R(\bar z)$ are conformal fields of dimension
$(1/2,0)$ and $(0,1)$ respectively, and $c$ is the ghost field.
For the $\Phi$ field to be a modulus field we must also have
\eqn\Viii{\vev{(V_{\Phi})^n}=0,}
in the zero momentum limit, \ie, $\Phi$ has no potential. An effective
Lagrangian for the $\Phi$ field may have a form
\eqn\iX{{\cal L}=\partial\Phi\partial\Phi^\dagger
+A\Phi\Phi^\dagger\partial\Phi\partial\Phi^\dagger+\cdots,}
(where the existence of the second term can be explicitly checked by evaluating
the correlator $\vev{V_{\Phi}V_{\Phi}V_{\Phi^\dagger}V_{\Phi^\dagger}}$).
{}From our experience with no-scale supergravity theories we know that the
above
Lagrangian is actually symmetric under a noncompact symmetry group
$SU(1,1)/U(1)$ \LN. (To realize this symmetry a field redefinition of $\Phi$ is
necessary \refs{\IB,\Ferr}.) The modular symmetry of interest is actually a
subgroup of the
above mentioned continuous symmetry. In string theories, as opposed to
no-scale supergravity theories, this continuous symmetry is broken
to the discrete subgroup of modular transformations in Eq. \III{}
by the higher order terms in the superpotential and higher derivative terms in
the effective action.

The K\"ahler potential reflecting the invariance of the theory under the
noncompact continous symmetry is
\eqn\Xi{K=-3\ln(T+\ov T).}
The transformation or the (non-holomorphic) field redefinition necessary to
show the equivalence between the effective Lagrangian of Eq. \iX\ and the one
obtained through \Xi\ is \refs{\IB,\thresh}
\eqn\Xii{\Phi={1-T\over1+\ov T}.}
The inclusion of the matter fields is straightforward. If, for a moment,
we just concentrate on the K\"ahler function $K$, in the string basis one has
\eqn\Xiii{K=\Phi\Phi^\dagger+A(\Phi\Phi^\dagger)^2+m'm^{'\dagger}
+B(m'm^{'\dagger})(\Phi\Phi^\dagger)+\cdots,}
where $m'$ is a matter field (in the string basis). On the other hand, in the
supergravity basis $K$ describes a sigma model on the coset space
$SU(1,N)/SU(N)\times U(1)$ with K\"ahler potential given by \EKN
\eqn\Xiv{K=-3\ln(T+\ov T-\sum_{i=1}^N m_im^\dagger_i).}
It is clear that in order for $K$ to be invariant (up to a K\"ahler
transformation) under the $PSL(2,{\bf Z})$ duality transformation, the fields
$m_i$ must transform with modular weight $-1$,
\eqn\Xv{m_i\to{e^{i\lambda(a,b,c,d)}\over icT+d}m_i,}
where $\lambda$ is a phase factor (called the multiplier system).
Note that the matter field $m'$ in the
string basis must also undergo a field redefinition for it to have the
simplified transformation properties given in Eq. \Xv\ \refs{\IB,\Ferr} (the
exact transformation \thresh\ is not critical for the present discussion).

Now we turn our attention to the main topic of this paper, the modular
transformation properties of the superpotential $W$. The point that the
detailed knowledge of $W$ and its transformation properties are critically
important in any serious analysis does not need to be belabored. The dichotomy
of the situation is that the detailed information about $W$ can only be
obtained in the string basis, whereas the symmetry properties of $W$ under
modular transformations are manifest only in the supergravity basis.

The invariance of the effective Lagrangian and consequently of $\cal G$ under
modular transformations implies that the superpotential $W$ must transform
as
\eqn\Xvi{W\to {1\over(icT+d)^3}W,}
\ie, with modular weight $-3$. For a trilinear term in the superpotential
(if all the matter fields in the theory have modular weight $-1$) the
modular covariance is automatic. It is clear that no higher order term in
the superpotential can be constructed which has modular weight $-3$. However,
if an explicit calculation (albeit in the string basis) {\it requires} a
quartic and/or higher order nonrenormalizable term to be present in the
superpotential, consistency of the theory will require that the quartic term
$W_4$ be of the form
\eqn\Xvii{W_4=m_1m_2m_3m_4\eta^2(T){\cal H}_4(T),}
where $\eta(T)$ is the Dedekind function of the first kind and has
transformation property
\eqn\Xviii{\eta(T)\to e^{i\pi/12}(icT+d)^{1/2}\eta(T)\qquad{\rm as}\quad
T\to {aT-ib\over icT+d},}
and ${\cal H}_4(T)$ is an arbitrary modular invariant function \Cvetic.
That is, the superpotential at the Nth order must be accompanied by appropriate
number of $\eta^2(T)$ powers to ensure that the overall invariance of $W$
is maintained,
\eqn\Xix{W=W_3+\eta^2(T){\cal H}_4(T)W_4+\eta^4(T){\cal H}_5(T)W_5+\cdots,}
where $W_N$ is the Nth order nonrenormalizable term and the ${\cal H}_N(T)$
are arbitrary modular invariant functions. As shown below,
generalization of the above construction to include twisted fields (matter
fields with modular weight $\not=-1$) and to include more than one type of
modulus field is straightforward.

When examined from the the vantage point of the string basis, the above
construction implies a remarkable structure. In terms of explicit
correlator calculations, the above statement entails that a nonzero value
of the quartic (or higher order) term in the superpotential necessitates the
existence of an infinite string of nonvanishing correlators, \ie,
\eqn\XX{\vev{m'_1m'_2m'_3m'_4}\not=0\quad\Rightarrow\quad
\vev{m'_1m'_2m'_3m'_4\Phi^n}\not=0,\quad {\rm for\ all\,}n.}
That is, a quartic (or higher order) term calculation in the string basis
will necessarily receive corrections due to the moduli fields. Note that
since the trilinear term is modular covariant by itself, no such correction
is necessary. Due to the arbitrariness of the modular function that may
appear in the superpotential in the supergravity basis (\ie, the ${\cal
H}_N(T)$
in Eq. \Xix) and due to the (non-holomorphic) field redefinition involved,
it is not always possible to compare specific numerical values obtained in
these
two different bases. However, a great deal of information about the structure
of the superpotential can still be obtained as we will show shortly.

In Ref. \STAB, an explicit calculation of the superpotential
for a class of models showed that the trilinear terms did not receive
higher order moduli-dependent corrections, \ie, they were stable.
Here we see that it is the modular invariance of the theory which is
responsible for the observed stability. Furthermore, this same symmetry
of the theory also dictates that the quartic and higher order
terms must receive corrections if the modulus field is away from its
canonical value (Eq. \Xii: $T\not=1\Rightarrow\Phi\not=0$).
\newsec{The free fermionic case}
The salient features of the free fermionic formulation of the heterotic
string (in four dimensions) \FFF\ include the existence of 18 real
two-dimensional
left-moving fermions $\chi^k,y^k,w^k\,(k=1,\ldots,6)$ transforming in the
adjoint representation of $SU(2)^6$. A model can be further divided into
`complex fermions' or `real fermions' model depending on whether or not
all two-dimensional fermions can be paired up to give bosonic fields.
Bosonization of the fields
\eqn\XXi{{1\over\sqrt{2}}(\chi^k+i\chi^{k+1})=e^{iS_{k,k+1}},\,k=1,3,5}
plays a special role since
\eqn\XXii{i\partial_z(S_{12}+S_{34}+S_{56})=J(z),}
where $J(z)$ is the conserved $U(1)$ current of the N=2 worldsheet
supersymmetry
algebra \KLN. A vertex operator for the scalar component of a chiral superfield
in the canonical ghost picture is given by \KLN
\eqn\XXiii{V^b_{-1}=e^{-c}e^{i\alpha S_{12}}e^{i\beta S_{34}}e^{i\gamma S_{56}}
G e^{{i\over2}k\cdot X}e^{{i\over2}k\cdot\ov X},}
where $G$ is a conformal field of dimension $(h(G),1)$, with
$h(G)=(1-\alpha^2-\beta^2-\gamma^2)/2$,
$\alpha,\beta,\gamma\in\{0,\pm{1\over2},\pm1\}$, $c$ is the ghost field,
and $\alpha+\beta+\gamma=1$.

This class of models can be shown to have modular group of at least
$PSL(2,{\bf Z})^3$ and three associated moduli fields which will be denoted
by $T_I,\,I=1,2,3$. The modular weight of each matter field is related to its
charge under $\partial_zS_{k,k+1},\,k=1,3,5$. A consistent choice is \thresh
\eqn\XXiv{m_i\to m_i\prod_I(ic_IT_I+d_I)^{-w^i_I},}
where $w^i_I$ is the charge of the vertex operator of the scalar component of
the $m_i$ superfield under the current $\partial_zS_{2I-1,2I}$ (\ie, the
$\alpha,\beta,\gamma$ in Eq. \XXiii).
This choice automatically
assures that all the trilinear couplings obtained using the S-matrix approach
are modular invariant since the conservation of $U_J(1)$ charge requires that
the superpotential carries charges $(-1,-1,-1)$ under $(S_{12},S_{34},S_{56})$.
\foot{As $U_J(1)$ and hence $(S_{12},S_{34},S_{56})$ are $R$-type charges,
$\theta$ carries a nontrivial charge under these transformations. Thus, for
$\int d^2\theta\,W$ to be invariant, $W$ must carry a nonzero charge.} That is,
Eq. \Xvi\ is generalized to
\eqn\XXiva{W\to\prod_I(ic_IT_I+d_I)^{-1}\,W.}
In terms of supergravity fields the K\"ahler function is analogously
generalized to \refs{\IB,\Ferr,\thresh}
\eqn\XXv{K=\sum_I-\ln(T_I+\ov T_I)
+\sum_i\prod_I(T_I+\ov T_I)^{-w_I^i}m_im_i^\dagger+\cdots.}

As far as the trilinear terms are concerned, no term consistent with the
gauge symmetry and other symmetries present in the theory is prohibited or
constrained in any way due to modular symmetry. This follows from the
conservation of $U_J(1)$ charged mentioned above. By the same token it is
also clear that no quartic or higher order term will be trivially covariant
under the modular symmetry. Since each superfield has modular weight
$w=-\sum_I w_I=-1$, Nth order nonrenormalizable terms
will have weight $-N$. The overall modular weight of the superpotential
must add up to $-3$ (or $(-1,-1,-1)$ under each $PSL(2,{\bf Z})$ separately).
Hence, if the theory is expected to accommodate modular invariance, then
either the offending Nth order term is zero or it must be accompanied by a
coupling which would depend on the moduli fields $T_i$ and would cancel
the excess modular charge of $-N+3$.

{}From explicit S-matrix calculations we know that some of these
nonrenormalizable
terms are definitely nonzero, we conclude then that their coupling must be
such as to screen out the modular weight. The simplest case of one modulus
field was discussed above (see Eq. \XX).
For the case of more than one modulus field the previous statement becomes
more precipitious. Knowing the $w_I$ charges of the fields it is
straightforward
to determine the ``deficit" modular weight of a particular term. The ``deficit"
will dictate the form and argument of the modular function necessary to cast
the term under consideration into modular covariant form. In the string
basis this implies the existence of correlators of the type
\eqn\XXvii{\vev{m'_1m'_2\cdots m'_N\Phi^p_i\Phi^k_j}\not=0\quad\forall p,k,}
where $i,j$ depend on the modular ``deficit" charge.

We now show how the above general remarks apply to a specific string model
built in the free fermionic formulation, namely the flipped $SU(5)$ string
model
\revamp. The fields $\Phi_1$ and $\Phi_2$ (in the string basis) correspond
to the moduli fields $T_1$ and $T_2$ (in the supergravity basis). We defer
until
later the discussion on the third modulus field $T_3$ (which appears to be
related to the fields $\Phi_4$ and $\Phi_5$).
The whole set of quartic superpotential couplings and their respective
(nonvanishing) coefficients is known in this model \decisive. For instance,
the quartic term $c F_1\bar f_1\bar h_{45}\phi_1$ has modular weight
$(-1,-2,-1)$ and therefore its coefficient must contain
$\eta^2(T_2){\cal H'}(T_2)$ to make it modular covariant. This implies that
in the string basis the correlators
$\vev{F_1\bar f_1\bar h_{45}\phi_1\Phi^n_2}$, $\forall n$ should all be
generically nonvanishing, a fact that has been verified explicitly for small
$n$. The actual quartic coefficient will then be
$c(T_2)=c\eta^2(T_2){\cal H'}(T_2)/\eta^2(1){\cal H'}(1)$.
We have confirmed that all quartic terms present in the superpotential do
indeed receive higher order corrections due to the moduli fields as anticipated
by the `modular deficit' argument. A similar analysis for the quintic and
higher order terms also bears out the premise that all nonrenormalizable terms
are accompanied by a proper modular function such as to make the superpotential
modularly covariant.

The flipped $SU(5)$ model discussed here differs from a typical model obtained
in the free fermionic formulation in one important respect, namely the presence
of an anomalous $U_A(1)$ in the gauge group.
The model appears to only have $PSL(2,{\bf Z})^2$ symmetry instead of
$PSL(2,{\bf Z})^3$ expected in free fermionic models of this class since there
is no $T_3$ field that can be readily identified as a modulus field.
We attribute the possible non-existence of the ``modulus" field $T_3$
(related to $\Phi_4$ and $\Phi_5$ in the string basis) to the ill understood
anomalous $U_A(1)$ phenomenon. Although it is quite conceivable that since the
theory does have additional flat directions, the modular symmetry group may be
$g'\times PSL(2,{\bf Z})^2$. However, we will not address this possibility
here.
The cancellation of the $U_A(1)$ anomaly requires a `discrete' shift in the
string vacuum to a different nearby vacuum \DSW.
This mechanism, which albeit cancels the anomaly and restores the
supersymmetry,
spontaneously breaks the modular symmetry from $PSL(2,{\bf Z})^2$ to either
$PSL(2,{\bf Z})$ or a smaller subgroup thereof depending on the choice of
vacuum expectation values.

The phenomenological implications of the structure of the nonrenormalizable
terms dictated by modular invariance are self evident. For example, while
looking for quartic or higher order terms, one need not consider terms of the
type $\vev{m_1m_2m_3\Phi}$ since they would automatically be zero. Nth order
($N\ge4$) terms of the type $\vev{m_1\cdots m_N\Phi^p}$ are nonzero only if
$\vev{m_1\cdots m_N}\not=0$. The strength of this observation lies in the fact
that the terms of the type $\vev{m_1\cdots m_N\Phi^p}$ are typically not
forbidden by any other symmetry of the theory. In Ref. \conden\ this structure
of the nonrenormalizable terms was explicitly used to determine the specific
value of the gaugino condensate. One expects that such information about the
structure of the theory will prove invaluable in any detailed phenomenological
inquiry.
\newsec{Conclusions}
In conclusion, we have argued that the requirement of modular invariance
imposes a strong constraint on the Nth order nonrenormalizable terms that may
be nonvanishing. We have verified explicitly that for free fermionic models,
the
structure of the nonrenormalizable terms implied by modular invariance is
indeed present. The most useful phenomenological advantage of the above result
is that it allows one to examine the physical implications of a model under a
small shift of the moduli fields. We further argued that the effective action
obtained in the string basis (after a suitable field redefinition) can be
recast to manifestly display its invariance under modular transformations.

The complete effective action essentially contains three separate pieces.
The gauge and `kinetic energy' (K\"ahler function) part, nonperturbative
effects of strong gauge interactions, and the superpotential. The modular
properties of the gauge and kinetic energy part were examined in
\refs{\Kap,\Cvetic,\anomaly,\Lacaze,\thresh} in the guise of `threshold
corrections', whereas the modular properties of the nonperturbative gauge
interaction terms have been known for some time. Here we complete the picture
by explicitly showing how the nonrenormalizable part of the superpotential
transforms under the modular group. As expected, all three pieces alluded to
above orchestrate themselves to embody modular invariance in the effective
theory. Further study of modular invariance, especially in the context of
its possible phenomenological implications will indubitably prove to be
very useful.
\bigskip
\noindent{\it Acknowledgments}: This work has been supported in
part by DOE grant DE-FG05-91-ER-40633.
\listrefssave
\bye